\documentclass[prc,aps,twocolumn,preprintnumbers,
amsmath,amssymb,showpacs,floatfix]{revtex4}
\usepackage{amsmath}
\usepackage{graphicx}
\usepackage{floatflt,epsfig}
\usepackage{lscape}
\newcommand\be{\begin{equation}} 
\newcommand\ee{\end{equation}} 
\newcommand\bea{\begin{eqnarray}} 
\newcommand\eea{\end{eqnarray}} 
\newcommand\pni{\par\noindent}
\begin{document}
\title{Event generator to construct cross sections for the\\
       multiphonon excitation of a set of collective vibrational modes}
\author{C.H. Dasso}
\affiliation{Departamento de F\'{i}sica At\'omica, Molecular y Nuclear,
Universidad de Sevilla, Apdo 1065, E-41080 Sevilla, Spain}
\author{E.G. Lanza}
\affiliation{I.N.F.N.-Catania and Dipartimento di Fisica e Astronomia,
Universit\'a di Catania,  Via S. Sofia 67, I-95123 Catania, Italy }
\author{H.M. Sofia}
\affiliation{Comisi\'on Nacional de Energ\'\i a At\'omica, 
 Buenos Aires, Argentina and CONICET} 

\date{\today}
\begin{abstract}
  The construction of differential cross sections as a function of
  excitation energy for systems with a collection of low- and
  high-lying intrinsic vibrational modes has been attempted in the
  past.  A prescription is proposed that simplifies the implementation
  of such calculation schemes with a remarkable reduction in
  computational time.
\end{abstract}
\pacs{21.10.Re, 21.10.Dr, 21.60.Ev, 21.60.Jz}
\maketitle

\section{Introduction}
\pni The possibility of exciting collective vibrational modes in heavy
ion reactions has captivated the attention of researchers in the field
for several decades \cite{BM}.  Of especial interest has been the
challenge of extending the set of participant states (i.e.~the open
channels) well beyond the familiar low-lying surface modes and into
the range of the nuclear giant resonances of the lowest
multipolarities.  Because of adiabatic considerations these processes
required, at the bombarding energies available during the 1970's and
80's, very short effective collision times.  Only the sharp
exponential drop of the nuclear radial formfactors was able to
recreate these conditions at that time, a fact that was noted (and
exploited) in the early studies of Deep Inelastic Collisions
\cite{bdw1,bdw2}.
\par
Nowadays the accelerators provide much higher beam energies and thus
the population of giant resonances can also be mediated by the Coulomb
excitation mechanism.  Under these circumstances the long range of the
Coulomb-coupling matrix elements forces us to incorporate --~in the
theoretical analyses of the process~-- a considerably larger number of
impact parameters or partial waves.  This, naturally, increases the
chances for exciting the high-energy part of the nuclear response.
\par 
Historically, the multiple-phonon excitation of high-lying modes was
actively promoted for being the source of characteristic structures in
the experimental distribution of cross sections as a function of
excitation energy\cite{dgr}. The study of these patterns became, in
turn, a quite convenient source of information for learning about the
actual features of the giant resonances (energies of the modes,
widths, strengths, anharmonicity, etc).
\par
This link motivated an important body of recent theoretical
work~\cite{lan,alt} where microscopic calculations for the structural
aspects of the giant modes have been combined with standard reaction
formalisms to yield concrete predictions for the shape of the
distributions $d\sigma/dE$.  Unfortunately, as it was stated earlier,
a much larger number of impact parameters are now needed to compute
accurately the nuclear and Coulomb components of the excitation
processes.  This, together with the fact that there exists a distinct
possibility of exciting a multiple number of collective phonons, has
resulted into rather complicated and time-consuming coupled-channel
schemes.
\par
Upon close inspection of the results, however, one realizes that for
practically the entire range of relevant partial waves the excitation
probabilities are very small.  It is also possible to conclude that,
in leading order, the different modes can be considered as being
independent from each other.  Or, in other words, that one can ignore
terms in the hamiltonian that involve simultaneously the coordinates
of two or more collective variables $\alpha_{\lambda\mu}$.  Notice
that this is {\em not} the same as claiming that one works within the
perturbation limit; we have already stressed the relevance of
multi-step events.  Multiphonon processes may occur as the excitation
of the same mode (two, three phonons) or as the simultaneous
excitation of two or more different modes \cite{lan}.
\par
Considering the harmonic modes as being essentially uncoupled to each
other has a significant practical advantage; it can be exploited to
design an event generator that allows for a much simpler, and yet
accurate, method for constructing the differential cross sections
$d\sigma/dE$.  We shall describe this idea in detail in the following
sections.
\par
Such an approximation scheme is of course bound to fail for the very
central impact parameters.  However, these partial waves are unlikely
to participate directly in the population of the inelastic channels
explicitly taken into account. Within the semiclassical formalism
their contribution is, in fact, strongly suppressed by the global
absorption associated with the imaginary part of the optical
potential.  To take into account properly the grazing impact
parameters may perhaps require a complete procedure as the one
exploited by the authors of~\cite{lan}.  Notice that this would be
necessary, at worst, for only a narrow window of impact parameters;
for most of the relevant range (extending up to hundreds of fermis)
the method we are proposing in this paper is appropriate and it is
precisely here that the major savings of computational time can be
achieved.
\par
We elaborate further on these considerations in Sect.~II, where we
also recall a previous work that proves quite essential to the
development of our proposal.  In Sect.~III we explain what is truly
specific about the technique for generating the folding of
probabilities that we need.  In Sect.~IV we take, as an example, the
reaction $^{40}$Ar~+~$^{208}$Pb at 40~MeV per nucleon and include two
low-lying and three high-lying resonances.  The purpose is here mainly
to show that the function $d\sigma/dE(E)$ indeed reflects correctly
the assumed input to the problem.  We reserve Sect.~V for a brief
summary of the contribution and some closing remarks.
\section{Brief Background}
\pni A few years back we developed a general formalism for the
excitation of a single collective vibrational mode~\cite{dflv}.  We
begin the presentation mentioning this reference because it proved
quite essential to motivate the present contribution.  Exploiting
well-tested approximations we were able in \cite{dflv} to propose a
semiclassical prescription to estimate the dynamical effects
associated with a spreading width $\Gamma$ of the mode.  Modifications
in the predicted cross sections due to the presence of an eventual
anharmonicity, manifested by an apparent ratio of state energies
$\varepsilon$ 
\be 
\nu= {{\varepsilon({\rm 2~phonon~state})} \over
  {\varepsilon({\rm 1~phonon~state})}} ~\ne~2~, 
\ee
\noindent
were also investigated.  We should mention here that the development
of this program --~aiming mostly to the description of single- and
double-phonon giant resonances~-- was done within the framework of
perturbation theory.
\par
Without entering into details in the implementation of
ref.\,\cite{dflv} let us briefly recall what was the input of the
calculation scheme and what could be obtained as a result.  Given a
single collective vibrational mode of multipolarity $\lambda$, energy
$\hbar\omega_\lambda$, width $\Gamma_\lambda$ and anharmonicity
$\nu_{\lambda}$ a run of the program densely sampling impact
parameters $\rho$ belonging to the interval
[$\rho_{min}$,~$\rho_{max}$] generated the total differential cross
section $d\sigma/dE_\lambda(E)$.  The energies of the transitions
$0\to$~1~phonon $1\to$~2~phonon, affected as they are by the
anharmonicity factors and the spreading widths, introduced interesting
dynamical consequences which were the main object of investigation in
\cite{dflv}.
%
%
\begin{figure}
\includegraphics[angle=0, width=0.90\columnwidth]{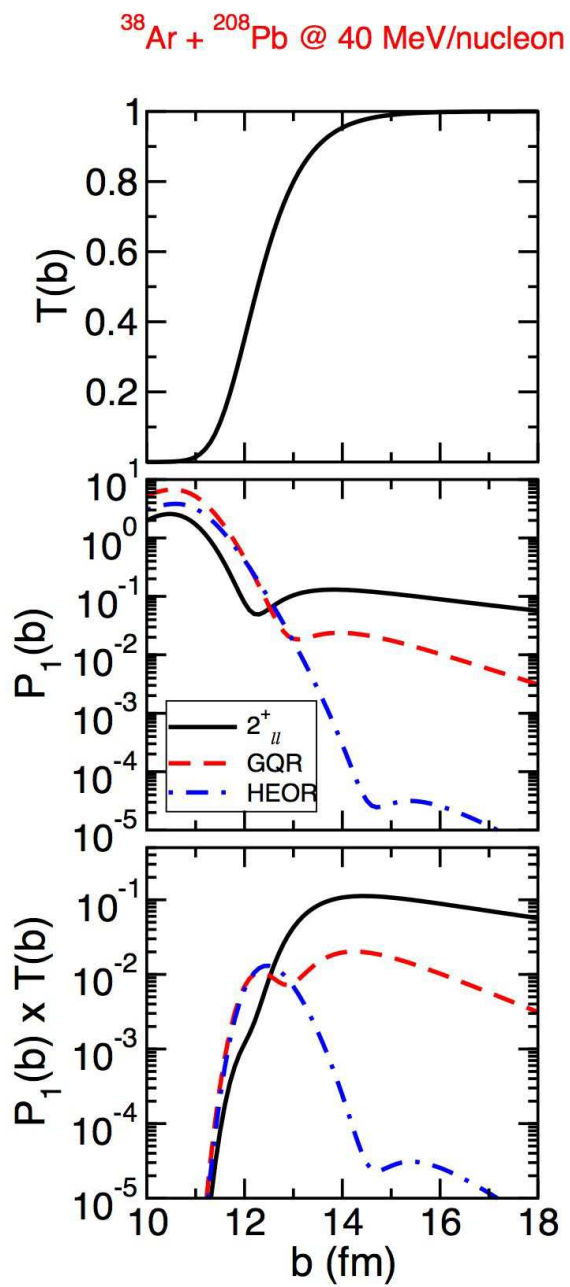}
\caption{(Color online) 
 Treatment of the absorption. The figure on the top shows the behaviour of 
 T(b) as function of the impact parameter b. In the middle frame, the 
 excitation probability of a single phonon is plotted as function of b 
 for the three states reported in the legend. The functions obtained as a
 product of the two previous quantities is plotted in the figure at the bottom. 
 }
\end{figure}
%
%
\par
To construct the differential cross sections $d\sigma/dE\,(E)$ it was
necessary to define a procedure to distribute the probabilities of
inelastic transitions for the one- and two-phonon vibrational states
over the relevant excitation energy ranges.  This prescription is
quite analogous (except for its generalization from one mode to
several modes) to the one we later use for our event generator.  We
can thus defer its presentation to the next Section.
\par
The effect of the absorption was taken into account in \cite{dflv} by
means of a multiplicative depletion factor that rapidly falls to zero
as the overlap between the reacting nuclei increases for the lower
impact parameters.  This is --~following standard practice~--
constructed from an integral along the trajectory $r(t)$,

\be
 T(b) = \exp \Big\{-{\frac{2} {\hbar}}
 \int_{-\infty}^{+\infty} W(r(t')~dt'\Big\}
\ee
\noindent
of the imaginary part of the optical potential, $W$ (see for instance,
\cite{bro} and refs. quoted therein).  For the current application the
absorptive component was chosen following the prescription of
refs.~\cite{dflv,for}.

The projection into the subspace of explicitly considered channels is
now able to reverse the tendency of the second-order amplitudes to
yield too large probabilities for impact parameters at or inside the
grazing distance.  Notice that this uncomfortable situation would not
become obvious when solving for coupled-channel amplitudes (even if
the numerics may be equally absurd) because of a prescribed
conservation of the norm by the integration algorithm.
\par
The interplay between these contrasting effects can be appreciated in
Fig.~1 for the reaction $^{38}$Ar + $^{208}$Pb at 40~MeV per
nucleon. The figure shows the attenuation factor that defines what
fraction of the contribution of a given impact parameter is actually
retained in the inelastic channels. The top frame shows the
transmission coefficient $T(b)$ as a function of impact parameter and
puts in evidence that the transition from $T(b)\approx 1$ to
$T(b)\approx 0$ indeed occurs over an interval $\Delta b$ that spans
only a couple of fermis. The middle frame displays the excitation
probabilities as a function of the impact parameter in three
situations.  These are, namely, the low-lying quadrupole mode (full
curve), the giant quadrupole resonance (GQR, dashed curve) and a high
energy octupole resonance (HEOR, dotted curve).  The lower frame
corresponds to the actual distribution $d\sigma/db$, constructed from
the information displayed in the two plots above.
\section{Formalism}
\pni 
We consider a pair of reacting heavy ions that accumulate a
number $N$ of intrinsic surface vibrational modes.  Each one is
characterized by their excitation energy $\hbar\omega_i$,
multipolarity $\lambda_i$, strength $\beta_i$ and a width $\Gamma_i$.
That is,

\be
 \hbar\omega_i,~~~\lambda_i,~~~\beta_i,~~~,\Gamma_i
 ~~~~~~~~~{\rm with}~~~~1\le i\le N~.
\ee

\noindent
We proceed immediately to define two separate groups of these modes:
high- and low-lying modes.  The mean reason for establishing the
subdivision has to do with the way people in the field have
traditionally dealt with their widths.  The differentiation is
actually an old story that dates back to the 70's and the use of
surface vibrational models to describe specific dynamical features of
Deep Inelastic Reactions \cite{bdw1,bdw2}.  It aims to reflect two
clear experimental facts:
\begin{itemize}
\item {Low-lying modes (excitation energies $\hbar\omega_\lambda\,<$
    5-6~MeV) are such that at the zero- and one-phonon level are sharp
    and display no width.  Clearly the energy range quoted above is
    only qualitative.  The modes we have in mind are those, for
    example, known --~in the harmonic oscillator terminology for even
    multipolarities~-- as $\Delta N$=0 excitations.  At the two-phonon
    level they show a spread which is mainly associated with the
    anharmonicity of the mode.  Let us be a bit more explicit.
    Suppose we have a low-lying quadrupole mode with
    $\hbar\omega_{\lambda=2}\approx4$ MeV.  At about double that
    excitation energy it is found a multiplet of states $0^+$, $2^+$,
    $4^+$ spanning an interval of energy that we shall call $\Delta
    E_{\lambda=2}$.  Typically this quantity has an order of magnitude
    of about 0.5 MeV.\hfill\break Finally, the last piece of
    experimental evidence that one wishes to incorporate in the
    formalism is that there are practically no known three-phonon
    states associated with low-lying modes.  This can be formally done
    by assuming that, at the three-phonon level, the mode assumes a
    width that equals the separation energy
    $\hbar\omega_\lambda$.\hfill\break All of these features are best
    implemented by ascribing to the mode an energy-dependent width
    (see below).}
\item {High-lying modes (i.e. giant resonances).  In this situation
    the zero-phonon state is taken to be sharp, while at the
    one-phonon level the state displays the well-known spreading width
    $\Gamma_\lambda$.  The distribution of the excitation amplitudes
    to higher levels is achieved by a straightforward folding (see
    below).}
\end{itemize}
\par
The prescription that emerges from the two items listed above may
appear, at first, difficult to grasp.  However it has led to practical
conclusions in the treatment of Deep Inelastic Collisions that are in
very good agreement with the experimental evidence.  Obviously one
could come up with different but somewhat equivalent operating
procedures.  It is easier and reasonable, however, to adhere to this
established practice since the details of its implementation have
already been described and tested in the literature (see for example
ref.\cite{bdw2}).  A practical reminder of the spreading prescription
as it is applied for low- and high-lying modes is summarized in
Table~1.
%
%
%
\begin {table}[b]
\begin{center}
\begin{tabular}{|c|c|c|c|c|}
\hline
 $n_\lambda$& ~0~ & ~1~      & 2               & 3               \\ \hline
 high-lying & ~0~ &~$\Gamma$~&$\sqrt{2} \Gamma$&$\sqrt{3} \Gamma$\\ \hline
 low-lying  & ~0~ & ~0~      & $\Delta E$       &   $E$           \\ \hline
\end{tabular}
\end{center}
\caption {\label{tab} Width prescription as it is used in the construction
 of the cross section shown in figures 2 and 3.}\smallskip\end{table}
%
%
\vglue 0.3truecm
\par
%
%
\begin{figure}
\includegraphics[angle=0, width=0.94\columnwidth]{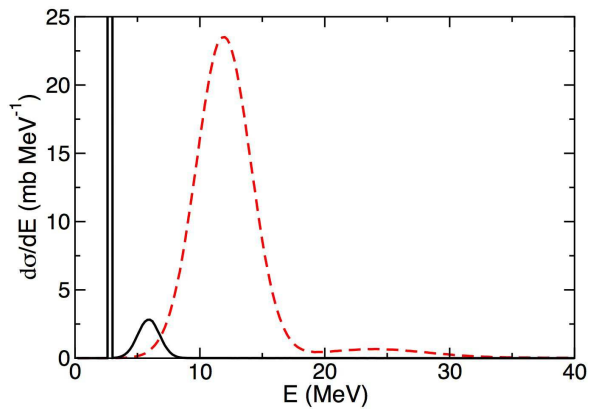}  
\caption{
  \label{width2} (Color online) Width prescription as applied to low-
  and high-lying modes.  The distributions of the total differential
  cross section $d\sigma/dE(E)$ is constructed as an example for two
  separate quadrupole modes with $\hbar\omega$=3 and 12~MeV.  In this
  figure we can appreciate the different prescriptions for low- and a
  high-lying modes summarized in Table I. The giant quadrupole state
  ($\Gamma(n=1)$= 5~MeV) clearly shows the second phonon with a wide
  and increasing width.  The low-lying mode ($\Delta E$= 1 MeV)
  displays, on the other hand, a much narrower structure at the
  two-phonon level and has the three phonon structure practically
  washed out (not visible in the figure).  A large number of events
  and an impact parameter range of 100~fm was used for this
  illustration.  }
\end{figure}
%
%
\par
The event-generator works in practice just like any other similar
device in a wide variety of physics subjects.  We take a given impact
parameter and consider a very large number of possible ``events'',
$N_{ev}$.  In every instance one generates the probability of
occurrence of each independent mode by ``throwing a Poisson dice'' $N$
times, in agreement with an average number of phonons $\langle n_i
\rangle$.  This means that the random number generator is designed to
return a number of phonons $n_i$ for each of the independent modes
consistent with the law 
\be 
P(n_i)={ { \langle n_i \rangle^{n_i}}
  \over {\langle n_i \rangle\,!} } ~\exp\{{-\langle n_i \rangle}\}~.
\ee
\par
The differential cross section we search for is to be constructed for
a large range of relevant impact parameters $[\rho_1,\rho_2]$, that
are sampled with a uniform interval $\Delta\rho$.  The average number
of phonons for the $i^{\rm th}$ independent collective surface mode is
previously calculated, for each impact parameter $\rho_k$, using the
formalism developed in ref.~\cite{dflv}.  These figures are collected
in an auxiliary data-file that has the structure 
\bea
\label{datarho}
 \rho_1, & \langle n_{i=1}\rangle, ... , \langle n_{i=N}\rangle \nonumber \\ 
   &... \nonumber \\
 \rho_k, &\langle n_{i=1}\rangle, ... , \langle n_{i=N}\rangle\\
    &... \nonumber \\ 
 \rho_2, &\langle n_{i=1}\rangle. ... , \langle n_{i=N}\rangle ~.\nonumber
\eea
\par
The smaller impact parameter, $\rho_1$ should be such that any
possible contribution from it is definitely eliminated by the
absorption.  Similarly, one should verify that contributions from
$\rho > \rho_2$ can also be neglected.
\par
Adding to this prepared data-set the characteristic information that
specifies the different vibrational modes $i$ one is ready to run the
event generator and construct $d\sigma(E)/dE$.
\par
For each impact parameter in the file (\ref{datarho}) the probability
assigned to the current event is, naturally, 
\be 
P =\prod_{i=1}^N~P(ni)\,.
 \label{labpro}
\ee
\par\noindent
It is in the energy scale that we have to be careful with the
character of low-lying vibrational state or giant resonance of the
particular mode $i$.  Following the prescription summarized in Table~1
we assign a spread $\Gamma_i$ to the contribution to the excitation
energy, $\epsilon_i$, of this mode. This quantity is thus defined as

\be
 \epsilon_i = n_i\,\hbar\omega_i\,+\,{\cal G}(0,\Sigma_i)~,
\ee
\noindent
where ${\cal G}(0,\Sigma_i)$ is a random number obtained from a normal
distribution with zero centroid and standard deviation
$\Sigma_i\approx\Gamma_i/2.3$.  Slightly different prescriptions could
once more be obtained by replacing ${\cal G}$ by a similar type of
random number generator, but these choices are not of much consequence
at the level of approximation we have chosen to maintain.
\par
The total excitation energy for the collection of $N$ independent
modes is then simply given by 
\be 
E\,=\,\sum_{i=1}^N~\epsilon_i~, 
\ee
\noindent
absissa to which --~in a properly designed histogram~-- we accumulate
a weighted version of the probability previously given in
eq.\,(\ref{labpro}).  The proper units for the differential cross
section as a function of energy are obtained by multiplying that
number by $(20\pi\rho_k\Delta\rho)/(\Delta E N_{ev})$, where $\rho_k$
is the impact parameter sampled at this moment, and $\Delta\rho$,
$\Delta E$ are the extents of the impact parameter mesh and the energy
mesh respectively.  The final result is then given in mb/MeV and a
drawing of this histogram represents the predicted distribution of
cross section $d\sigma(E)/dE$.
%
%
\begin {table}[b]
\begin{center}
\begin{tabular}{|c|c|c|}
\hline
 states& events generator & ref.\cite{lan} calculations         \\ \hline
 $3^-$            & 1.10                 & 1.05                 \\ \hline
 GQR              & 58.7                 & 58.6                 \\ \hline
 $3^- \times 3^-$ & 0.11 $\times 10^{-3}$& 0.10 $\times 10^{-3}$\\ \hline
 $GQR \times GQR$ & 0.14 $\times 10^{-1}$& 0.14 $\times 10^{-3}$\\ \hline
 \end{tabular}
\end{center}
\caption {\label{tab2} Total excitation cross section (in mb)
  calculated with the events generator method (second column) and with
  the method of ref.\cite{lan} (third column) for the states shown in
  the first column.}\smallskip\end{table}
%
%
\par
To test the validity of our method, a comparison with a more
sophisticated method like the one of ref.~\cite{lan} is in order. In
this approach one starts with a Hartree-Fock plus Random Phase
Approximation calculation in order to identify the most collective
one-phonon states. For each of these chosen states one calculates
their transition densities and the corresponding form factors. These
are used in a semiclassical coupled channel scheme to determine the
excitation probabilities for all the possible one-, two- and
three-phonon states that one can construct.
\par 
In Table II we compare some results for the two approaches. The
calculations are performed for the system $^{40}$Ca + $^{208}$Pb at 50
MeV per nucleon. We take a simple example where only two one-phonon
states are considered as input: the low-lying $3^-$ state (E=4.9 MeV)
and the GQR (E=16.9 MeV). The range of impact parameter used in the
calculations (15-100 fm) corresponds to the peripheral region where
the Coulomb interaction yields the most important contribution. This
is also the region where the excitation probability distribution are
of a Poisson type. The results of the two methods are very close.
\par
So, one can envisage a calculation performed in two steps: Make use of
the method of ref.~\cite{lan} in the inner region where the nuclear
interaction plays an important role and then use our novel approach in
the peripheral region for large impact parameters which is the most
time consuming part.
\par
Finally, we would like to stress that in the case one wants to take
into account the contribution of anharmonicities, our method does not
apply.
%
%
\begin{figure}[b]
\includegraphics[angle=0, width=0.94\columnwidth]{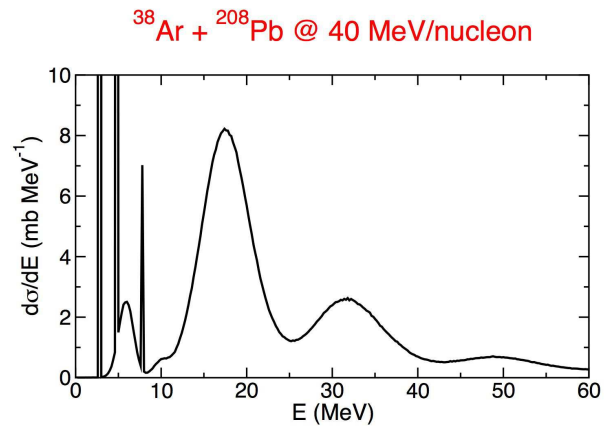}  
\caption{
\label{dens}
Constructed distributions of the total differential cross section
$d\sigma/dE(E)$ for the reaction $^{38}$Ar + $^{208}$Pb at 40 MeV per
nucleon.  Some 10$^8$ events per impact parameter have been generated
and their contributions accumulated to construct this figure.  The
calculation time is however a few minutes CPU.  The example is
described in detail and the relevant parameters given in the text of
Sect.~IV.  }
\end{figure}
%
%
%
\section{Application}
\pni 
We now proceed to illustrate the possibilities of the event
generator with an application to the specific reaction $^{38}$Ar +
$^{208}$Pb at 40 MeV per nucleon. We take two low-lying modes (one
quadrupole, one octupole) with energies $\hbar\omega$=3 and 5 MeV
respectively, and a common value $\Delta E$=1 MeV. The deformation
parameter assumes for the low-lying modes a value $\beta$=0.1~. Three
giant resonances are then added; a dipole, a quadrupole and an
octupole mode with energies $\hbar\omega$=18, 17, and 31~MeV and
widths $\Gamma$=6, 6, and 8 MeV respectively.  The effective spread of
all these modes is found in all cases following the prescriptions
described in Table~1.  Accumulating the cross section for a range of
impact parameters from $\rho_{min}$=12~fm to $\rho_{max}$=100~fm in
steps of 0.5~fm we obtain the distribution shown in Fig.~3.
\par
The characteristics of the curve reflect the assumptions made and, in
addition, are reminiscent of what is obtained by the time-demanding
method of ref.~\cite{lan} in similar circumstances. Just to stress the
advantage of the proposed method let us note that one can save about
two orders of magnitude in computing time by using the event
generator.
\section{Concluding Remarks}
\pni 
The developments described and the results shown in this Report
were motivated by sheer necessity.  In fact, while the complete
calculations performed by Lanza {\em et al.}  are very important the
absolute times involved in the computation of cross sections by the
procedure described in ref.~\cite{lan} are quite large.  This becomes
even more critical when one takes into account that the practical
implementation of such a prescription involves a considerable amount
of leeway that can only be sorted out by trying alternative
calculations with equally acceptable sets of parameters.  That appears
to be the only sensible way to learn, by gaining experience, how the
various input numbers do indeed affect the calculated distributions of
cross section.  Even if a final presentation is contemplated with the
full procedure of \cite{lan} a number of previous calculations
exploiting the event generator would be --~no doubt~-- very convenient
to prepare the ground.
\par
We consider that the use of an event generator like the one described
in these pages is highly advisable for that class of problems that can
use suggestive results to judge the soundness of the answers they
provide.  Actually this should be done even {\rm before} embarking in
sophisticated schemes without the proper benefit of an educated
intuition.  
\vskip 1.0truecm 

\acknowledgements{ Support is acknowledged from the Spanish Ministry
  of Education and
  Science under project numbers FIS2005-01105, FPA2005-04460.\\
  Two of us E.G.L and H.M.S. want to acknowledge the kind hospitality
  of the Departamento de F\'{\i}sica At\'omica, Molecular y Nuclear de
  la Universidad de Sevilla, where this work has been done.}

\vskip 1.0truecm

\begin{thebibliography}{99}
\bibitem{BM}  Aa. Bohr and B. R. Mottelson, in {\em Nuclear Structure},
              Benjamin, New York, 1975 and references quoted herein.
\bibitem{dgr} H. Emling {\em Prog. Part. Nucl. Phys.} {\bf 33} 729 (1994);
              Ph. Chomaz and N. Frascaria {\em Phys. Rep.} {\bf 252} 275 (1995); 
              T. Aumann, P. Bortignon and H. Emling
              {\em Annu. Rev. Part. Sci.} {\bf 48} 351 (1998);
              C. A. Bertulani and V. Yu. Ponomarev 
              {\em Phys. Rep.} {\bf 321} 139 (1999).
\bibitem{bdw1} R. A. Broglia, C. H. Dasso and Aa. Winther, {\em Phys. Lett. B}
               {\bf 53} 301 (1974); {\em Phys. Lett. B}{\bf 61} 113 (1976).
\bibitem{bdw2} R. A. Broglia. C. H. Dasso and Aa. Winther, in 
               {\em Nuclear Structure and Heavy Ion Collisions}, eds.
               C. H. Dasso and R. A. Broglia, Aa. Winther, North Holland,
               Amsterdam, 1980
\bibitem{lan} E. G. Lanza, M. V. Andr\'es, F. Catara, Ph. Chomaz
              and C. Volpe, {\em Nucl. Phys. A} {\bf 613} 445 (1997);
              E. G. Lanza, F. Catara, M.V. Andr\'es, Ph. Chomaz, 
              M. Fallot and J. A. Scarpaci, {\em Phys. Rev. C} 
              {\bf 74} 064614 (2006).
\bibitem{alt} V. Y. Ponomarev et al., {\em Phys. Rev. Lett.}
               {\bf 72} 1168 (1994); V. Y. Ponomarev, P. F. Bortignon, 
               R. A. Broglia and V. V. Voronov, 
               {\em Phys. Rev. Lett.} {\bf 85} 1400 (2000);
               C. A. Bertulani, L. F. Canto, M. S. Hussein and 
               A. F. R. de Toledo Piza, {\em  Phys. Rev. C} {\bf 53} 334 (1996);
               B. V. Carlson, M. S. Hussein, A. F. R. de Toledo Piza and 
               L. F. Canto, {\em  Phys. Rev. C} {\bf 60} 014604 (1999);
               J. Z. Gu and H. A Weidenm\"uller, {\em Nucl. Phys.A}
               {\bf 690} 382 (2001). 
\bibitem{dflv} C. H. Dasso, L. Fortunato, E. G. Lanza and A. Vitturi, 
               {\em Nucl. Phys. A} {\bf 724} 85 (2003).
\bibitem{bro}  R. A. Broglia and Aa. Winther, {\em Heavy Ion Reactions}, 
               Addison-Wesley Publishing Company, 1991;
               R. A. Broglia, S. Landowne, R. A. Malfliet, V. Rostokin 
               and Aa. Winther, {\em Phys. Rep.} {\bf 11} 1 (1974).
\bibitem{for}  L. Fortunato, W. von Oertzen, H. M. Sofia and A. Vitturi,
               {\em Eur. Phys. J. A} {\bf 14} 37 (2002).
\end{thebibliography}
\end{document}